\documentclass[runningheads]{llncs}
\usepackage{graphicx}
\usepackage{booktabs}
\usepackage{tabularx}

\newcolumntype{C}{>{\centering\arraybackslash}X}
\newcolumntype{B}[1]{>{\centering\arraybackslash}p{#1}}

\begin{document}
\title{Advancing Brain Tumor Inpainting with Generative Models}
%
%
\author{Ruizhi Zhu\inst{1} \and
Xinru Zhang\inst{2} \and
Haowen Pang\inst{2} \and
Chundan Xu\inst{2} \and
Chuyang Ye\inst{2,*}
}
\authorrunning{R. Zhu et al.}
%
\institute{School of Information Engineering, Minzu University of China, Beijing, China \and
School of Integrated Circuits and Electronics, Beijing Institute of Technology, Beijing, China\\
$*$ Corresponding author: \email{chuyang.ye@bit.edu.cn}}
\maketitle              
\begin{abstract}
Synthesizing healthy brain scans from diseased brain scans offers a potential solution to address the limitations of general-purpose algorithms, such as tissue segmentation and brain extraction algorithms, which may not effectively handle diseased images. We consider this a 3D inpainting task and investigate the adaptation of 2D inpainting methods to meet the requirements of 3D \textit{magnetic resonance imaging}~(MRI) data. Our contributions encompass potential modifications tailored to MRI-specific needs, and we conducted evaluations of multiple inpainting techniques using the BraTS2023 Inpainting datasets to assess their efficacy and limitations.

\keywords{Brain tumor \and inpainting \and MRI.}
\end{abstract}
\section{Introduction}
Numerous algorithms for automated analysis of brain \textit{magnetic resonance imaging}~MRI images exist to aid clinical decision-making. It is, however, challenging to perform automatic image processing for patients with brain tumors as many algorithms are intended for analyzing healthy brains and may not effectively handle images with lesions. Examples of such algorithms include brain anatomy parsing, tissue segmentation, and brain extraction. We propose that synthesizing healthy brain scans from diseased brain scans could potentially address this challenge, and the problem is formulated as a 3D inpainting task~\cite{ref_1}.

Inpainting, a fundamental task in computer vision, has undergone significant advancements over the years. Its primary objective is to realistically fill in missing regions within 2D natural images, enabling various applications, including image restoration, object removal, and image completion. The computer vision community has developed numerous sophisticated algorithms that excel in addressing inpainting challenges in the context of two-dimensional images, achieving impressive results, and driving the state-of-the-art in this field.

Despite the remarkable progress made in 2D inpainting, the adaptation of these algorithms to the realm of 3D, especially in the context of medical imaging, remains an open problem. 
This paper aims to explore the challenges associated with 3D inpainting of MRI scans and investigate the feasibility of adapting existing 2D inpainting algorithms to the MRI data.
In this study, we put forth the following contributions:
\begin{itemize}
  \item [1)] 
  An exploration of potential modifications and adaptations of existing 2D inpainting techniques to better suit the requirements of 3D MRI data.       
  \item [2)]
  Presentation of experimental results and a comparative evaluation of different inpainting methods, assessing their efficacy and limitations on the BraTS2023 Inpainting datasets~\cite{ref_2}. 
\end{itemize}

\section{Methods}

Given images with brain tumors and tumor masks or with masks that simulates brain tumors, we input the masked image into an inpainting model, and the model is expected to output a healthy-looking brain where the tumor region is filled.
We selected three image synthesis models, namely pGAN~\cite{ref_3}, ResViT~\cite{ref_4}, and Palette~\cite{ref_5}, for our experiments. 
pGAN and ResViT are originally developed for brain MRI data, and we adapted the architecture of the Palette model to accommodate 3D data as input.

\subsection{pGAN}
\subsubsection{Preprocessing} Initially, we apply z-Score standardization to normalize the intensity. Subsequently, we slice the MRI data based on the slice dimension and save each slice as an individual 2D npy file.

\subsubsection{Network} pGAN~\cite{ref_3} is a conditional generative adversarial network composed of a generator $G$, a pre-trained VGG16 network, and a discriminator $D$. 
The generator $G$ learns to create target images that correspond to the healthy region. Simultaneously, the discriminator $D$ assesses the authenticity of the synthetic and real images. All subnetworks are trained jointly, where the generator $G$ aims to minimize pixel-wise, perceptual, and adversarial losses ($\mathcal L_{pix}$, $\mathcal L_{per}$, and $\mathcal L_{adv}$, respectively), and the discriminator $D$ aims to maximize the adversarial loss. The complete loss is 
\begin{equation}
\mathcal L_{pGAN} = \lambda_{pix}\mathcal L_{pix} + \lambda_{per}\mathcal L_{per} + \lambda_{adv}\mathcal L_{adv},
\end{equation}
where $\lambda_{pix}$, $\lambda_{per}$, and $\lambda_{adv}$ are the weights of the pixel-wise, perceptual, and adversarial losses, respectively.

\subsection{ResViT}
\subsubsection{Preprocessing} Preprocessing for ResViT is the same as that of pGAN.

\subsubsection{Network} ResViT~\cite{ref_4} is a conditional generative adversarial model built upon the self-attention Transformer architecture. The generator subnetwork adheres to the encoder-information bottleneck-decoder pathway, with the incorporation of a Transformer to explicitly leverage nonlocal contextual information. Meanwhile, the discriminator subnetwork is formulated based on a conditional PatchGAN architecture. The generator's bottleneck comprises a stack of novel \textit{aggregated residual transformer} (ART) blocks. Each ART block is organized as a cascade of a transformer module responsible for extracting hidden contextual features, along with a convolutional module that captures hidden local features from the input feature maps. The loss function incorporates both a pixel-wise L1 loss and an adversarial loss. The complete loss is 
\begin{equation}
\mathcal L_{ResViT} = \lambda_{pix}\mathcal L_{pix} + \lambda_{adv}\mathcal L_{adv},
\end{equation}
where $\lambda_{pix}$ and $\lambda_{adv}$ are the weights of the pixel-wise and adversarial losses, respectively.

\subsection{3D Palette}
\subsubsection{Preprocessing} To accommodate GPU memory limitations, we adopt the crop method utilized by the baseline model 3D Pix2Pix, cropping the image to a size of $96\times96\times96$ centered on the mask region. Additionally, the data normalization procedure aligns with that of the baseline model.

\subsubsection{Network} Palette~\cite{ref_5} is a conditional diffusion model that employs the Unet architecture~\cite{ref_6,ref_7}. The diffusion model transforms samples from a standard Gaussian distribution into samples from an empirical data distribution using an iterative denoising process. The conditional diffusion model tailors the denoising process based on the input signal. 
The model employs the L2 loss function to capture the distribution of the output.

\section{Results}
\subsection{Qualitative evaluation}
\begin{figure}
  \centering
  \rotatebox{90}{ 
    \begin{minipage}{\textheight} 
      \includegraphics[width=1\textwidth]{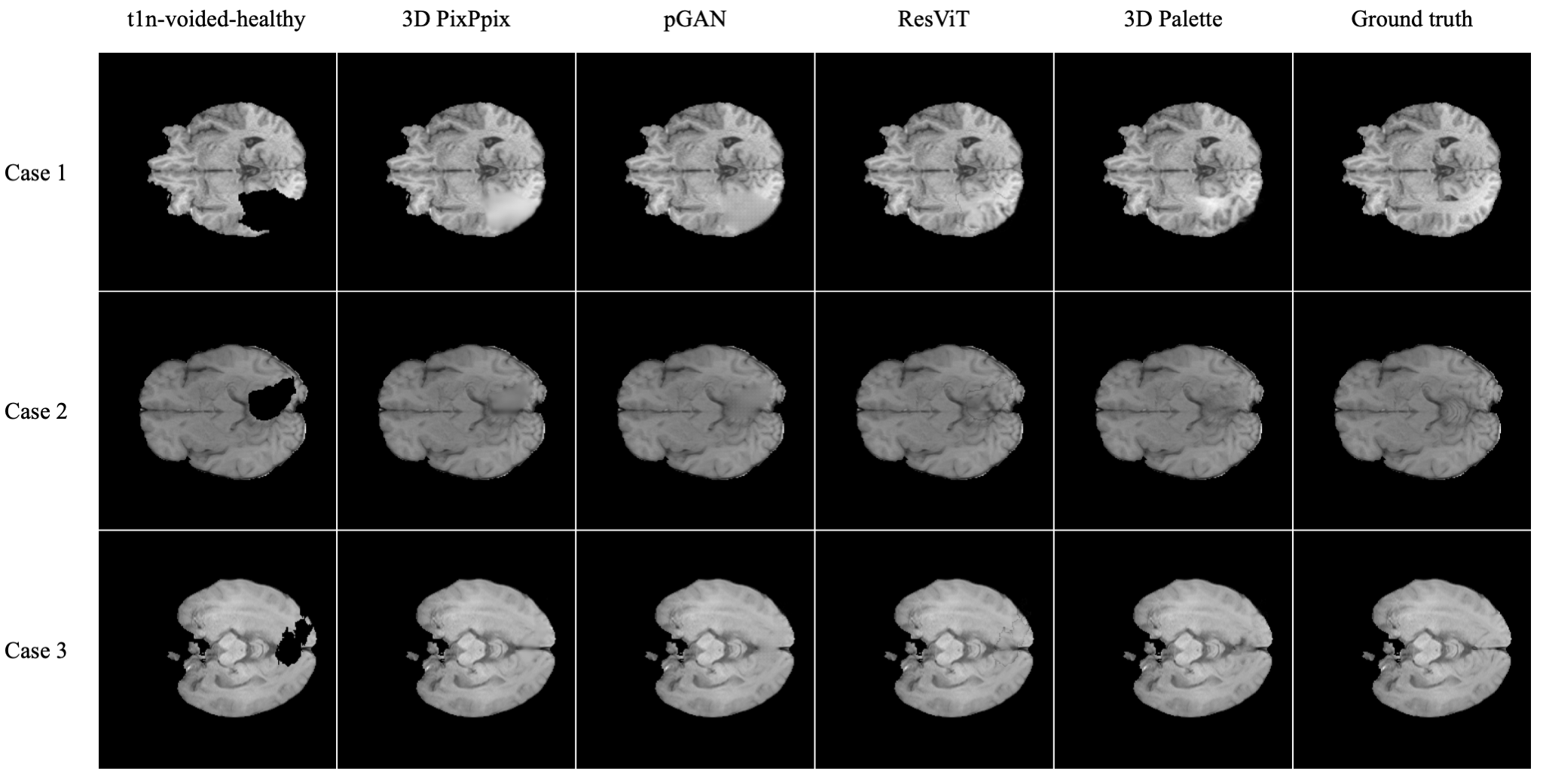} 
      \caption{Comparison of inpainting results for regions without tumors.}
      \label{fig1}
    \end{minipage}
  }
\end{figure}

\begin{figure}
  \centering
  \rotatebox{90}{ 
    \begin{minipage}{\textheight} 
      \includegraphics[width=1\textwidth]{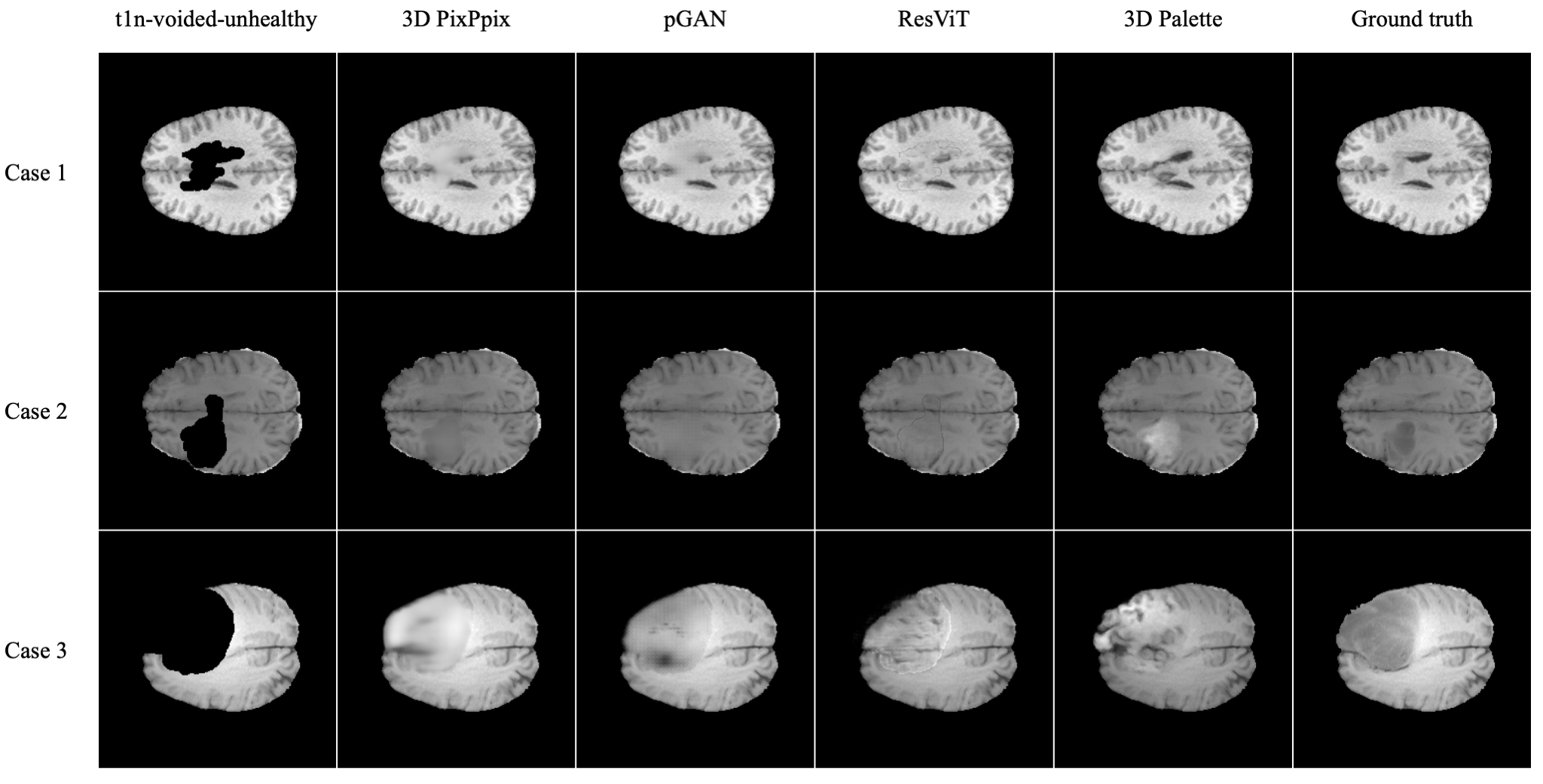} 
      \caption{Comparison of inpainting results for regions with tumors.}
      \label{fig2}
    \end{minipage}
  }
\end{figure}

We visually inspected the inpainting results of random samples for each of the three models and the baseline model. The cross-sectional views of the results are shown in Figs.~\ref{fig1} and \ref{fig2} for slices without and with brain tumors, respectively.

Both pGAN and ResViT use the same 2D slice as input. However, the results clearly indicate that ResViT generates output with substantially higher levels of detail. Likewise, when employing 3D MRI data as input, 3D Palette demonstrates greater potential than 3D pix2pix.

\subsection{Quantitative evaluation}
Following the evaluation guidelines of the MICCAI-BraTS2023 Inpainting Challenge, we quantitatively assessed our inpainting results with the \textit{structural similarity index measure} (SSIM)~\cite{ref_8}, \textit{peak-signal-to-noise-ratio} (PSNR), and \textit{mean-square-error} (MSE).

Considering the slow inference speed of 3D Palette, we conducted evaluations on only ten samples, whereas pGAN, ResViT, and the baseline model were evaluated on 250 samples in addition to the ten samples. The results are shown in Tables~\ref{tab1} and \ref{tab2}.

\begin{table}
  \centering
  \fontsize{10}{18}\selectfont
  \caption{Quantitative comparison (mean$\pm$standard deviation) of inpainting results on 250 samples.}
  \label{tab1}
  \begin{tabularx}{\textwidth}{B{2.1cm} *{3}{C}}
    \toprule
    \textbf{Model} & \textbf{SSIM} & \textbf{PSNR} & \textbf{MSE} \\
    \midrule 
    3D Pix2Pix & 0.8072 $\pm$ 0.1292 & 19.2986 $\pm$ 2.4886 & 0.0111 $\pm$ 0.0068 \\
    pGAN & 0.7679 $\pm$ 0.1531 & 18.0701 $\pm$ 2.7323 & 0.0159 $\pm$ 0.0132 \\
    ResViT & 0.7454 $\pm$ 0.1521 & 16.9280 $\pm$ 2.0234 & 0.0192 $\pm$ 0.0088 \\
    \bottomrule 
  \end{tabularx}
\end{table}

\begin{table}[!ht]
\centering
  \fontsize{10}{18}\selectfont
  \caption{Quantitative comparison (mean$\pm$standard deviation) of inpainting results on ten samples.}
  \label{tab2}
  \begin{tabularx}{\textwidth}{B{2.1cm} *{3}{C}}
    \toprule    
    \textbf{Model} & \textbf{SSIM} & \textbf{PSNR} & \textbf{MSE} \\
    \midrule 
    3D Pix2Pix & 0.7209 $\pm$ 0.0955 & 18.9827 $\pm$ 2.1998 & 0.0143 $\pm$ 0.0071 \\
    pGAN & 0.6746 $\pm$ 0.1044 & 17.5363 $\pm$ 2.9347 & 0.0219 $\pm$ 0.0141 \\
    ResViT & 0.6260 $\pm$ 0.1105 & 17.8071 $\pm$ 1.8430 & 0.0181 $\pm$ 0.0079 \\
    3D Palette & 0.6222 $\pm$ 0.1218 & 16.6225 $\pm$ 2.4153 & 0.0259 $\pm$ 0.0177 \\
    \bottomrule 
\end{tabularx}
\end{table}

\noindent An intriguing observation is that while ResViT produces results with more realistic details than pGAN, its performance measured by the metrics does not surpass that of pGAN.

%
%

\section{Discussion}
Based on the results of the qualitative evaluation, it is evident that for the inpainting tasks considered in this work, the models often experience a loss of detailed structure. 
For the inferior outcome metrics of ResViT in quantitative evaluation compared to 3D Pix2Pix, one possible reason could be related to slice continuity issues. 
The objective of our upcoming work is to modify the model structure of ResViT to enable its compatibility with 3D data. 
Furthermore, the combination of inpainting results from multiple models to achieve better inpainting quality may be explored in future. 
%
%
%

\begin{thebibliography}{8}
\bibitem{ref_1}
Kofler, Florian, et al. "The Brain Tumor Segmentation (BraTS) Challenge 2023: Local Synthesis of Healthy Brain Tissue via Inpainting." arXiv preprint arXiv:2305.08992 (2023).

\bibitem{ref_2}
A. Karargyris, R. Umeton, M.J. Sheller, A. Aristizabal, J. George, A. Wuest, S. Pati, et al. "Federated benchmarking of medical artificial intelligence with MedPerf". Nature Machine Intelligence. 5:799–810 (2023).

\bibitem{ref_3}
S. U. Dar, M. Yurt, L. Karacan, A. Erdem, E. Erdem and T. Çukur. "Image Synthesis in Multi-Contrast MRI With Conditional Generative Adversarial Networks." IEEE Transactions on Medical Imaging, vol. 38, no. 10, pp. 2375-2388 (2019)

\bibitem{ref_4}
O. Dalmaz, M. Yurt and T. Çukur. "ResViT: Residual Vision Transformers for Multimodal Medical Image Synthesis." IEEE Transactions on Medical Imaging, vol. 41, no. 10, pp. 2598-2614 (2022)

\bibitem{ref_5}
Chitwan Saharia, William Chan, Huiwen Chang, Chris Lee, Jonathan Ho, Tim Salimans, David Fleet, and Mohammad Norouzi. Palette: Image-to-Image Diffusion Models. In ACM SIGGRAPH 2022 Conference Proceedings (SIGGRAPH '22). Association for Computing Machinery, Article 15, 1–10. New York, NY, USA (2022). \doi{10.1145/3528233.3530757}

\bibitem{ref_6}
Ho, Jonathan, Ajay Jain, and Pieter Abbeel. "Denoising diffusion probabilistic models." Advances in neural information processing systems 33 (2020): 6840-6851.
\bibitem{ref_7}
Ronneberger, Olaf, Philipp Fischer, and Thomas Brox. "U-net: Convolutional networks for biomedical image segmentation." Medical Image Computing and Computer-Assisted Intervention–MICCAI 2015: 18th International Conference, Munich, Germany, October 5-9, 2015, Proceedings, Part III 18. Springer International Publishing (2015)

\bibitem{ref_8}
Z. Wang, A. C. Bovik, H. R. Sheikh, and E. P. Simoncelli. "Image quality assessment: from error visibility to structural similarity." IEEE transactions on image processing, vol. 13, no. 4, pp. 600–612 (2004)
\end{thebibliography}
%

\end{document}